%% file: main.tex
\documentclass[conference]{IEEEtran}
\IEEEoverridecommandlockouts
\usepackage{cite}
\usepackage{amsmath,amssymb,amsfonts}
\usepackage{algorithmic}
\usepackage{graphicx}
\usepackage{textcomp}
\usepackage{xcolor}
\def\BibTeX{{\rm B\kern-.05em{\sc i\kern-.025em b}\kern-.08em
    T\kern-.1667em\lower.7ex\hbox{E}\kern-.125emX}}
\begin{document}

\title{
Dynamic Difficulty Adjustment in Virtual Reality Exergames through Experience-driven Procedural Content Generation \\
}

\author{\IEEEauthorblockN{Tobias Huber}
\IEEEauthorblockA{\textit{University of Augsburg} \\
Augsburg, Germany \\
tobias.huber@uni-a.de}
\and
\IEEEauthorblockN{Silvan Mertes}
\IEEEauthorblockA{\textit{University of Augsburg} \\
Augsburg, Germany \\
silvan.mertes@uni-a.de}
\and
\IEEEauthorblockN{Stanislava Rangelova}
\IEEEauthorblockA{\textit{University of Augsburg} \\
Augsburg, Germany \\
stanislava.rangelova@protonmail.com}
\and
\IEEEauthorblockN{Simon Flutura}
\IEEEauthorblockA{\textit{University of Augsburg} \\
Augsburg, Germany \\
simon@flutura.de}
\and
\IEEEauthorblockN{Elisabeth André}
\IEEEauthorblockA{\textit{University of Augsburg} \\
Augsburg, Germany\\
andre@informatik.uni-augsburg.de}
}

\maketitle

\input{abstract}


\input{introduction}
\input{relatedwork}
\input{approach}
\input{evaluation}
\input{discussion}
\input{conclusion}

\section*{Acknowledgment}
This work presents and discusses results in the context of the research project ForDigitHealth.
The project is part of the Bavarian Research Association on Healthy Use of Digital Technologies and Media (ForDigitHealth), funded by the Bavarian Ministry of Science and Arts.
We thank our students Peter Fefelow, Nikolai Glaab, David Makowski, Luitpold Reiser, Alexander Renk, Rusmin Spahic, Sebastian Spolwind, Leon Wöhrl and Dennis Zürn for helping us to implement the prototype.
We thank Patrick Dohle and Stefan Künzell for helping us to design the physical exercises.

\bibliographystyle{IEEEtran}
\bibliography{IEEEabrv,literature}
\end{document}

%% file: abstract.tex
\begin{abstract}
Virtual Reality (VR) games that feature physical activities have been shown to increase players' motivation to do physical exercise.
However, for such exercises to have a positive healthcare effect, they have to be repeated several times a week.
To maintain player motivation over longer periods of time, games often employ \emph{Dynamic Difficulty Adjustment} (DDA) to adapt the game's challenge according to the player's capabilities.
For exercise games, this is mostly done by tuning specific in-game parameters like the speed of objects.
In this work, we propose to use experience-driven  \emph{Procedural Content Generation} for DDA in VR exercise games by procedurally generating levels that match the player's current capabilities. 
Not only finetuning specific parameters but creating completely new levels has the potential to decrease repetition over longer time periods and allows for the simultaneous adaptation of the cognitive and physical challenge of the exergame. 

As a proof-of-concept, we implement an initial prototype in which the player must traverse a maze that includes several exercise rooms, whereby the generation of the maze is realized by a neural network.
Passing those exercise rooms requires the player to perform physical activities.
To match the player's capabilities, we use Deep Reinforcement Learning to adjust the structure of the maze and to decide which exercise rooms to include in the maze.
We evaluate our prototype in an exploratory user study utilizing both biodata and subjective questionnaires.

\end{abstract}

%% file: introduction.tex
\section{Introduction}
Working and leisure conditions in today's society are shifting further from physical exertion to purely digital activities.
Especially with the ongoing COVID-19 pandemic and the associated restrictions to physical leisure activities, people do not achieve the recommended levels of physical activity.
The back in particular is at risk for postural damage due to monotonous and sedentary work.
One way to tackle this problem is to use Virtual Reality (VR) games that encourage physical exercise.
The use of VR exercise games (often called \emph{exergames}) has been shown to increase the motivation to do physical activity for workers in sedentary occupations \cite{yoo2020,costa2019virtual}.

In order for exercises to show a positive effect on the users' health, they should be done multiple times per week \cite{haskell2007physical}.
However, periodically repeating the same tasks often gets boring.
In order to keep players motivated over a longer period of time, we propose to combine two methods commonly used to tackle problems of repetition and boredom: \emph{Procedural Content Generation (PCG)} and \emph{Dynamic Difficulty Adjustment (DDA)}. 
DDA keeps players motivated by matching the game's challenge to the players' skill level.
When done correctly, this allows players to enter a state of flow between anxiety and boredom \cite{czikszentmihalyi1990flow,csikszentmihalyi2014toward}. 
For exergames, DDA has the additional benefit of adjusting the difficulty of the exercises such that they provide efficient physical training without overburdening the player \cite{Yoo2017}.
Incorporating PCG into exergames to create visually different levels has been shown to reduce repetition between different play sessions \cite{pezzera2019}.

\begin{figure}
    \centering
    \includegraphics[width=1\linewidth]{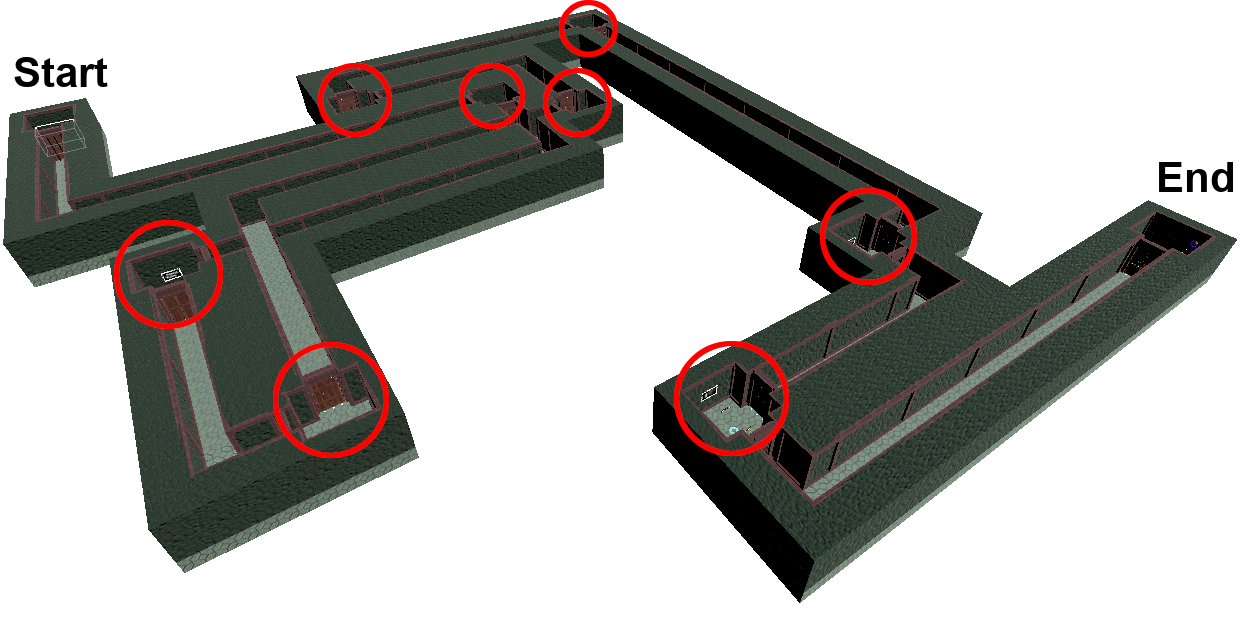}
    \caption{An exemplary maze generated by our prototype.
    Players have to traverse the maze while completing each exercise room they come across.
    During the game, players cannot see over the walls of the maze.
    The exercise rooms are marked by red circles.}
    \label{fig:ingame}
\end{figure}

In this work, we present a prototype of a VR exergame that utilizes PCG for DDA by creating game levels whose difficulty matches the player's capabilities.
In our prototype, the player has to traverse a maze that includes exercise rooms that have to be completed when the player wants to pass them.
As an exemplary use case, we select several exercises to prevent lower back pain.
The mazes are procedurally generated by a neural network that is trained with a Deep Reinforcement Learning (DRL) algorithm in order to adapt the difficulty of the generated mazes to the player.
An example for such a maze is shown in Fig. \ref{fig:ingame}.
The difficulty of the maze is mainly influenced by two factors: 1) the physical exertion during the exercise rooms and 2) the complexity of the maze.
More complex mazes are more difficult to traverse and players might have to repeat an exercise room several times.
Thus, adjusting the structure of the maze and the difficulty of the exercise rooms contained within the maze allows for implicit control of the physical and cognitive effort that the user has to put in.
Such a combination of physical and cognitive effort is often used in commercial exergames, such as \emph{Beat Saber\footnote{https://www.beatsaber.com/}}, one of the best-selling exergames.
In Beat Saber, the player must physically hit an incoming sequence of virtual blocks. 
The difficulty is determined by the complexity of the sequence (cognitive effort) and the time in which the user must hit the blocks (physical effort).
This combination prevents Beat Saber from becoming boring, as players can choose sequences that suit their cognitive and physical abilities.
However, Beat Saber only uses predefined sequences and does not dynamically adjust the difficulty.

The fact that the required physical effort and the complexity of the generated levels both have to be adapted to the player's capabilities makes exergames more challenging for DDA than traditional games.
If the current level is too complex but requires a fitting amount of physical effort, then the DDA algorithm should only adjust the complexity but not the physical challenge and vice versa.
As far as we are aware, our prototype is the first approach to explore the use of PCG for DDA in exergames, allowing for simultaneous adaption of the required cognitive and physical effort. 
Furthermore, it is the first prototype to show the feasibility of RL for DDA in a VR-based exergame.

%% file: relatedwork.tex
\section{Related Work}

\paragraph{VR Exergames increase physical activity} Using immersive VR games to motivate players to exercise has been explored since 2011 when Finkelstein et. al \cite{finkelstein2011} introduced the game Astrojumper. 
Here, players had to avoid incoming asteroids by moving their bodies.
While Astrojumper utilized a three-wall stereoscopic projected display, Charoensook et al. \cite{charoensook2019} showed that VR exergames using a Head-Mounted Display (HMD) can also increase players' heart rates during play sessions.
More recently, Yoo et al. \cite{yoo2020} allowed workers in a sedentary workplace to play a range of commercial VR games during work breaks over the duration of eight weeks and measured their exertion through questionnaires and a wearable heart rate monitor.
Their results show that the VR games motivated the workers over a longer period of time to be physically active.

\paragraph{Dynamic Difficulty Adjustment (DDA)}
To keep players engaged over several months, it is important that the games do not become boring when the players eventually get used to them. 
On the flip side, new players should not be overwhelmed with the challenges of the game.
When both the capabilities of the player and the challenges of the game are properly balanced, players enter a flow-state, feeling a deep sense of enjoyment \cite{czikszentmihalyi1990flow,csikszentmihalyi2014toward}.
For this reason, most single-player games continuously increase the difficulty during the course of the game. 
Such a predetermined difficulty increase, however, can never perfectly fit the different learning speeds of all players.
This is especially true when the difficulty is linked to physical exercises.
DDA tries to tackle this problem by adjusting the difficulty during gameplay based on the player's capabilities \cite{hunicke2005,zohaib2018}.

A common method for DDA is to apply Reinforcement Learning (RL). 
The basic idea of RL is that an agent interacts with an environment in order to maximize the accumulated reward given by a reward function.
By incorporating the player's performance into the action selection \cite{andrade2005} or the reward function \cite{pagalyte2020}, RL can be used to train AI opponents that play on the same level as the player.
Instead of training non-player characters, other approaches use RL to fine-tune specific in-game parameters such as speed and size of objects that directly influence the difficulty of the player's task\cite{sekhavat2017}.

\paragraph{Procedural Content Generation (PCG)}
In games, PCG refers to the autonomous generation of game content through algorithmic means \cite{yannakakis2018artificial}.
This is often used to increase replay value by creating vast amounts of different content without the need for more and more human designers and artists.
Such PCG systems in commercial games mostly do not take player behavior into account\cite{yannakakis2018artificial}.
For exergames, Pezzera et al. \cite{pezzera2019} showed that using PCG to create visually different levels without adapting to the player can already reduce repetition and therefore increase player motivation.

This work focuses on experience-driven PCG systems that are used for DDA by procedurally generating game levels that match the player's capabilities \cite{jennings2010polymorph}. 
Similar to our approach, Shaker et al. \cite{shaker2010} train a neural network to generate levels for a platform game based on the player's performance.
Others utilize Bayesian optimization \cite{gonzalez-duque2020} and evolutionary algorithms \cite{wheat2015} to create fitting levels procedurally.
For mazes in particular, van der Linden et al. \cite{van2013designing} propose graph
grammars to allow game designers to generate mazes of a specific difficulty levels.
In contrast to our approach, the aforementioned methods did not create content for exergames. 
The difficulty of their levels was mainly based on the complexity of the levels and did not have to adapt to the physical exertion of the player.
Balancing the cognitive and physical difficulty presents an additional challenge for a combination of PCG and DDA in exergames.

\paragraph{DDA in Exergames}
Despite the promising results of DDA for motivating players, there are only few VR exergames that use DDA and so far those games only tune small amounts of in-game parameters based on heuristics \cite{finkelstein2011,Yoo2017}.
In this work, we propose to use a combination of RL and procedural generation for DDA in VR exergames, based on the promising results of those approaches in the DDA literature.

For non-VR exergames, DDA is mostly used for rehabilitation games, where patients with varying degrees of impairments have very different requirements for the exergame and where the patients' capabilities might change drastically during recovery.
Besides increasing physical activity, rehabilitation is one of the main applications for exergames \cite{costa2019virtual}.
Many rehabilitation exergames focus on post-stroke rehabilitation \cite{burke09stroke_rehab,burke2010designing,alankus2010,hocine2011,sekhavat2017}.
Here, regular exercise can help to improve mobility in affected body parts.
However, other medical conditions, like Parkinson's, were also explored \cite{smeddinck2013,costa2019virtual,pezzera2020dynamic}.
The DDA approaches in many of those rehabilitation exergames adjust a small number of specific parameters in the game (e.g. the speed of in-game objects) according to heuristics based on the player's performance \cite{burke09stroke_rehab,burke2010designing,alankus2010,hocine2011,pezzera2020dynamic}.
Other rehabilitation exergames adapted similar in-game parameters through fuzzy systems \cite{pezzera2020dynamic}, evolutionary algorithms \cite{andrade2016} and RL agents \cite{sekhavat2017}.

Instead of adjusting specific in-game parameters, this work presents the first prototype for an exergame that uses DDA to procedurally generate in-game levels that match the player's capabilities. 
By doing so, we open the possibility to adjust the required physical and cognitive efforts simultaneously.

%% file: approach.tex
\section{Approach}
\label{sec:approach}

To show the feasibility of a DDA system that procedurally generates game levels adapting to the capabilities of the player in a VR exergame, we created a first prototype. 
In this prototype, the player has to traverse a procedurally generated maze that includes several exercise rooms (see Fig. \ref{fig:ingame}).
The cognitive challenge of the maze comes from the fact that players cannot see over the walls of the maze. 
Therefore, they must explore different paths in the maze and remember which paths they have already chosen. 
The physical challenge is given by the exercise rooms.
To pass through these rooms, the players must perform physical activities that are designed to prevent back pain (see section \ref{sec:exercise_rooms}).
At the end of the maze, the player rates how difficult and exhausting the maze was on a combined 5-point Likert scale (1-not at all difficult to 5-extremely difficult).
Based on this rating, the difficulty of the next maze is adjusted (see section \ref{sec:maze_generation}).
In this way, the difficulty of the generated mazes adjusts according to the player's training progress.
When the current maze was too easy, the next maze will get harder, and if the current maze was too challenging, the next one will be easier.

\subsection{The Exercise Rooms}
\label{sec:exercise_rooms}

\begin{figure}
    \begin{small}
    \centering
    \begin{minipage}[t]{0.2\linewidth}
        \includegraphics[width=\linewidth]{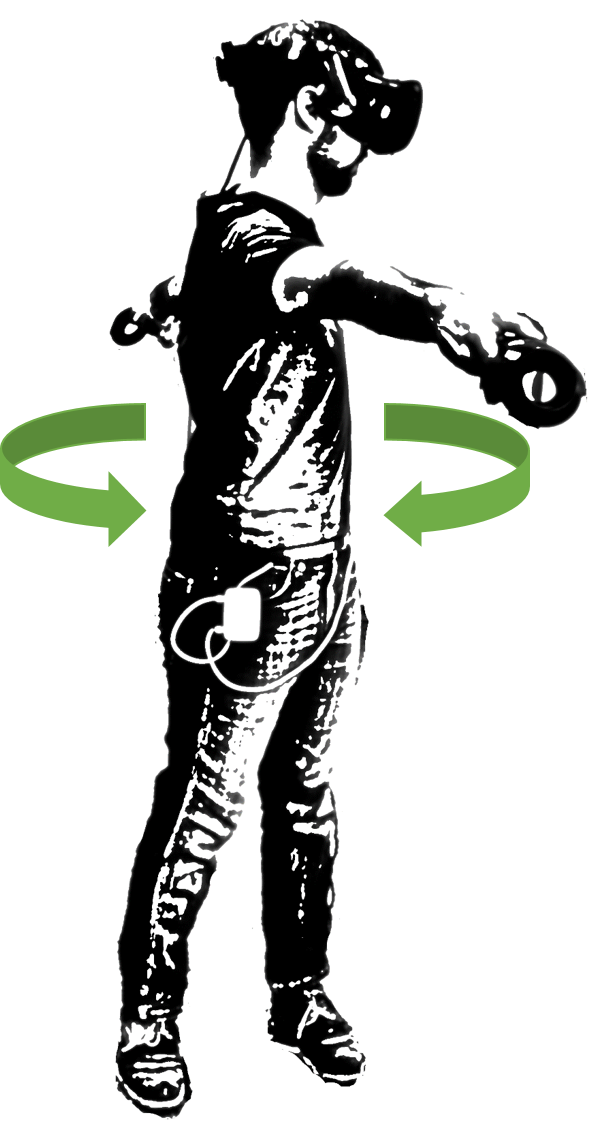}
        \parbox[t][0.5\baselineskip]{\linewidth}{\centering
        a) Rotation}
    \end{minipage}
    \hspace{0.05\linewidth}
    \begin{minipage}[t]{0.3\linewidth}
        \includegraphics[width=\linewidth]{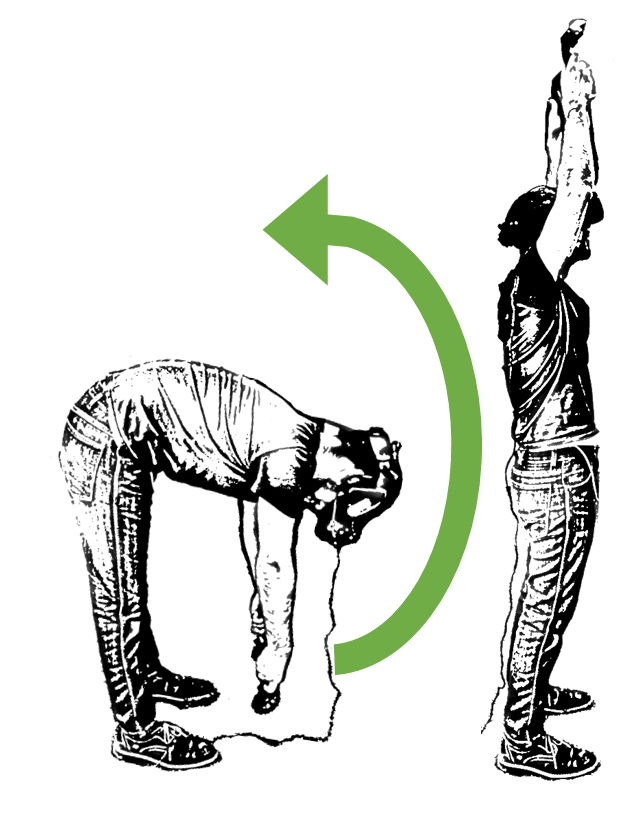}
        \parbox[t][0.5\baselineskip]{\linewidth}{\centering
        b) Torso Bend}
    \end{minipage}
    \hfill
    \begin{minipage}[t]{0.3\linewidth}
         \includegraphics[width=\linewidth]{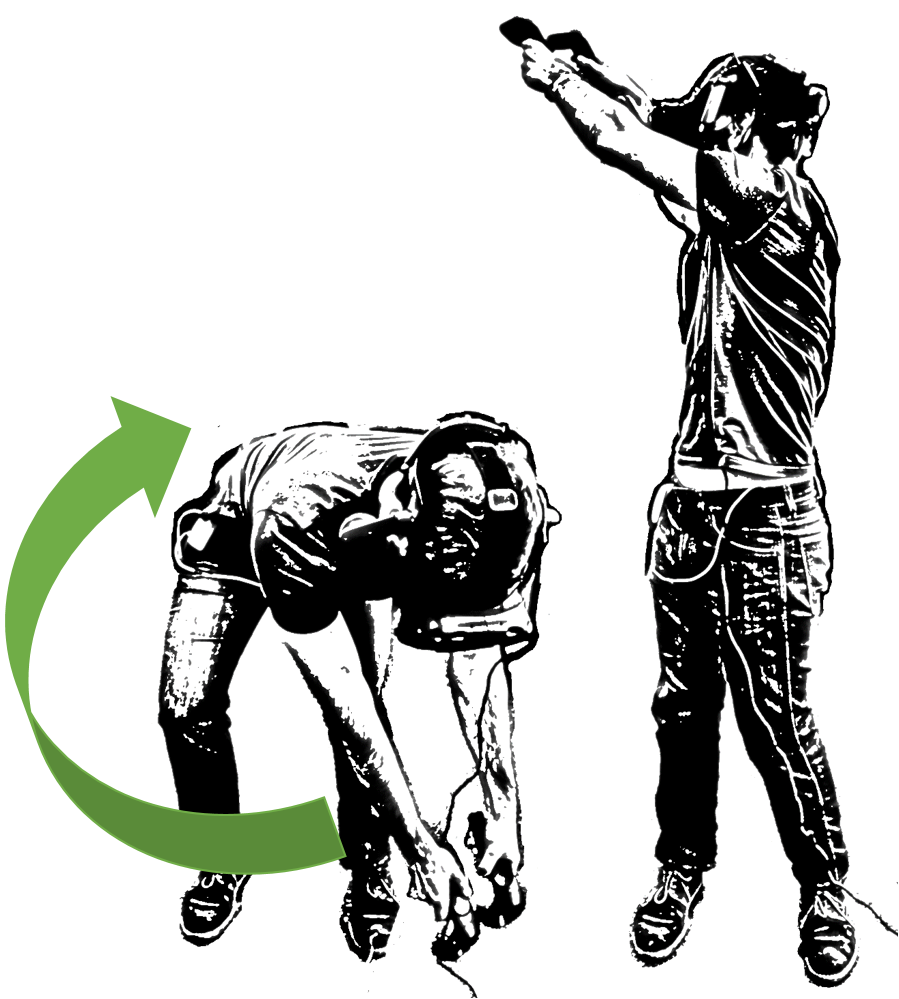}
         \parbox[t][0.5\baselineskip]{\linewidth}{\centering
        c) Bend and Stretch}
    \end{minipage}
    \end{small}
    \caption{The three exercises we implemented in our prototype VR exergame.}
    \label{fig:excercises}
\end{figure}

As an exemplary use case, we chose exercises to prevent back pain.
Since we want our prototype to be usable with standard commercial VR setups (we used the HTC VIVE Pro\footnote{https://www.vive.com/us/product/vive-pro/}) there are two main restrictions to the physical exercises we can use in the game. 
First, since it is very hard to lay down or get up while wearing an HMD and holding controllers in each hand, exercises that involve laying down are not feasible.
Second, the exercises should not depend on specific foot movements, since those can not be tracked with a basic VR setup.
Based on feedback from colleagues at the Institute of Sport Science at the University of Augsburg, 
we implemented the following three exercises as described in \cite{buskies2003ruckenfitness}:
\begin{itemize}
    \item \emph{Upper body rotation} (Fig. \ref{fig:excercises} a). 
     Participants must hold on to two in-game bars at shoulder height and move them left and right by rotating the upper body in a smooth motion.
    \item \emph{Forward torso bend} (Fig. \ref{fig:excercises} b).
    Players have to bend their torso forward, with legs extended and upper body as straight as possible, until they are able to grab an in-game bar laying on the ground.
    Subsequently, they have to slowly straighten their body and stretch upwards, moving the in-game bar slightly behind their back.
    \item \emph{Bending and stretching with torso rotation} (Fig. \ref{fig:excercises} c).
    From a hip-width stance, the players have to bend their legs and upper body as far as possible and turn the upper body to the left in order to pick up an in-game item.
    Then they have to stretch the upper body upwards and turn to the right in order to place the item on a platform.
    The exercise is repeated with alternating starting sides.
\end{itemize}

To verify that the exercises are done correctly, the game tracks the position of the controllers in the player's hands and checks whether they follow a predefined path describing the correct motion. 
We created variants of the rooms with different levels of exertion by varying the number of required repetitions of the physical exercise.

\subsection{Maze generation}
\label{sec:maze_generation}
\begin{figure*}
    \centering
    \includegraphics[width=1\linewidth]{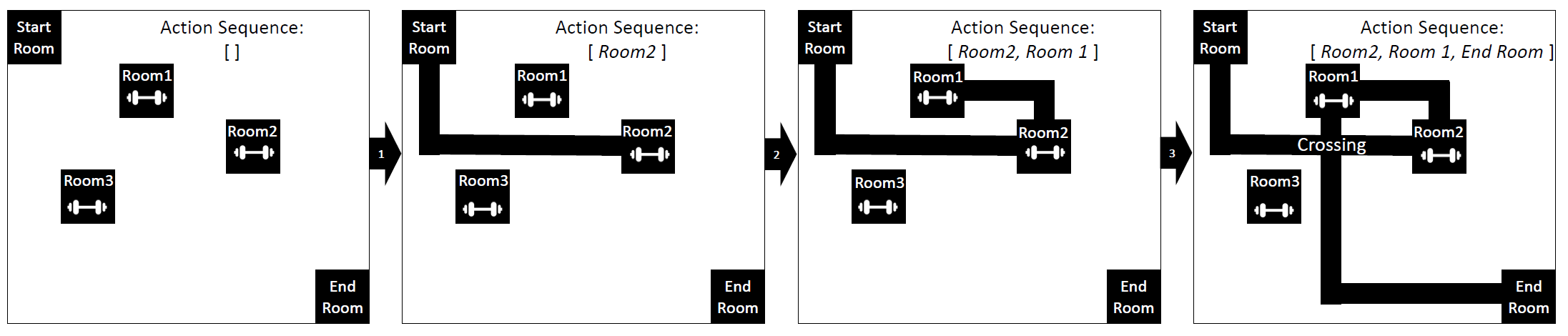}
    \caption{Exemplary sequence of our maze generation process. By iteratively placing connections to new rooms, a maze structure is created.}
    \label{fig:generation_steps}
\end{figure*}
\begin{figure}
    \centering
    \includegraphics[width=1\linewidth]{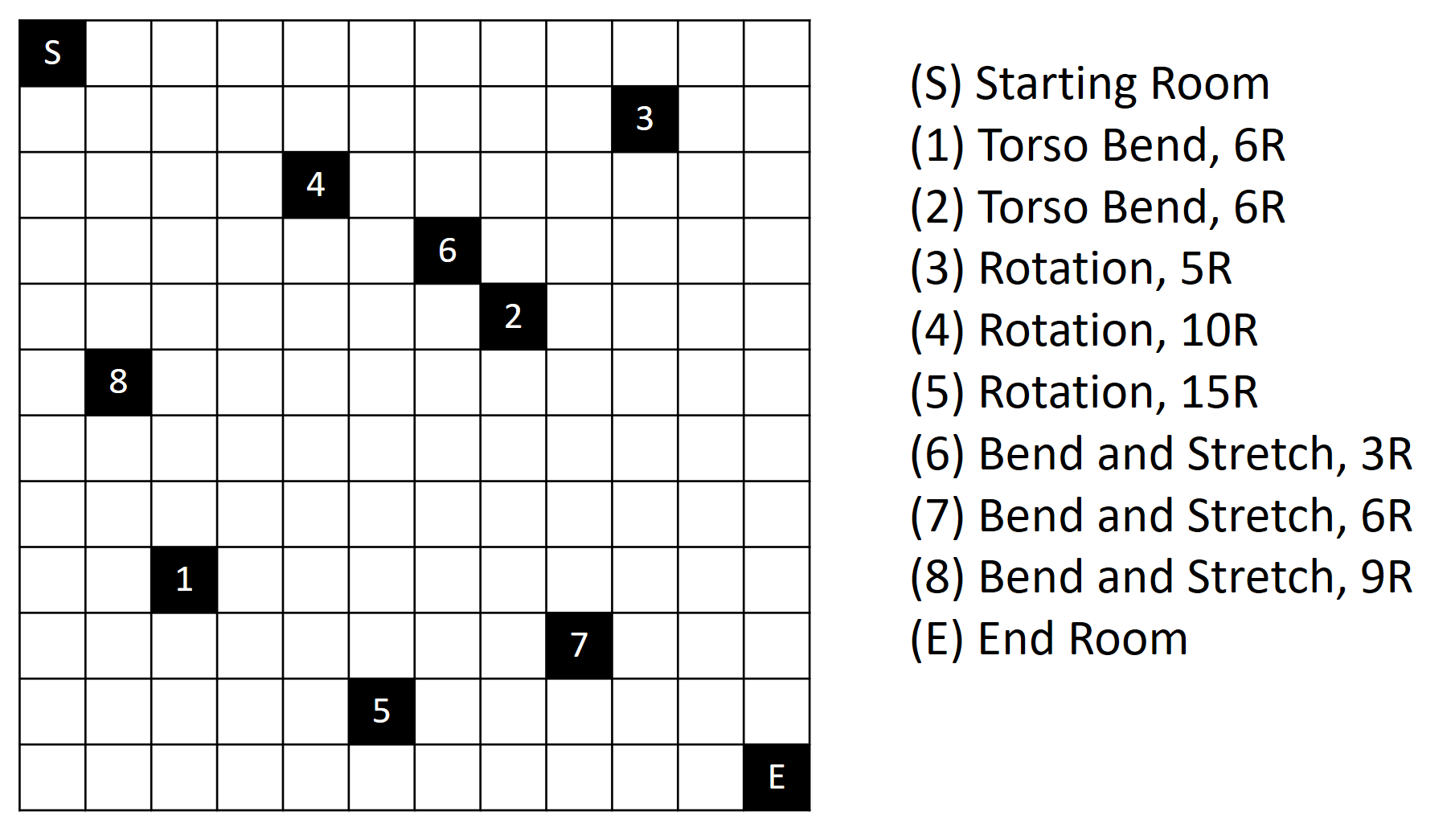}
    \caption{The grid that was used in our application. Eight exercise rooms were placed, containing the three different back pain exercises with varying numbers of repetitions (denoted with R in this figure).}
    \label{fig:grid}
\end{figure}
In order to enhance the level generation with the ability to adapt the difficulty of consecutively generated mazes to the individual player, we designed an adaptive level generation system.
The goal of that system is to create different maze structures that fit the player's needs with respect to cognitive and physical effort.
Thus, two major factors are responsible for a sufficient maze structure:
\begin{itemize}
    \item The structure of corridors has to be solvable in a way that the player does neither feel mentally overwhelmed by its complexity nor bored by too simple structures.
    \item The player should have to exert a reasonable degree of physical effort. Thus, exercise rooms have to appear in an appropriate frequency and difficulty while traversing through the maze.
\end{itemize}
RL has proven its ability to adapt the difficulty of games.
However, traditional RL algorithms suffer under exploding state spaces when dealing with complex relations like the structure of a maze.
Since deep learning is more suited to model complex relations, we decided to build the maze generation system based on DRL.
As most established DRL algorithms share the fact that the required amount of training effort grows substantially when dealing with high-dimensional input state spaces, we do not train the algorithm to create mazes from scratch.
Instead, we use a procedural approach similar to the PCGRL approach recently proposed by \cite{Khalifa2020}.
We designed a fixed grid of different exercise rooms, where each room consists of an exercise with a predefined difficulty level (see section \ref{sec:exercise_rooms}).
Fig. \ref{fig:grid} illustrates the final room grid that we used in our experiments. As can be seen, the position of the rooms was chosen in a way that the same exercises, although with different difficulties, are not positioned in close vicinity to each other. 
Further, to prevent the algorithm from adding two rooms at the same step, the exercise rooms were placed such that a single interconnection between any two rooms never results in a third room being crossed.

 In order to create a maze structure, we use DRL to connect the exercise rooms with corridors. 
 To this end, the DRL algorithm successively connects two rooms, whereby different sequences of connected rooms result in different interconnection structures. Further, many room connection sequences result in crossings between the generated corridors, leading to a variety of different maze instances. By learning the sequence of the room connections, the DRL algorithm is equipped with the ability to implicitly generate mazes of different difficulty. 
 See Fig. \ref{fig:generation_steps} for a simplified visualization of our generation process.

\subsection{Modeling the DRL problem:}
To allow the DRL algorithm to solve the learning problem stated above, we modeled it as a Markov decision process.

Since the goal of the DRL algorithm is to build a maze that fits the desired difficulty level, we use the player's difficulty rating at the end of each maze for the reward.
For every generated maze, the difference between the desired difficulty level (in our case $3$) and the actual player rating is given as negative reward to the DRL. 

As mentioned above, the DRL algorithm works iteratively, i.e., in every step, either a new interconnection between exercise rooms is made, or the final interconnection to the end room is made. Thus, the action space $A$ of the DRL algorithm is defined as follows: 
\begin{equation}\begin{aligned}
A = \{New \,connection \,to \,Exercise\, Room \,1, \\
New \,connection \,to \,Exercise\, Room \,2,\\
...,\\
New\, connection \,to\, Exercise\, Room\, n,\\
Connection\, to\, End \,Room\}
\end{aligned}
\end{equation}
Note that this implies that not every room is necessarily incorporated in the final maze.
This is a key factor to being able to adapt to the player's physical needs, as a reduced number of exercise rooms results in less physical effort.

Since every action generates a new interconnection, intermediate mazes are created. Thus, in every step, the maze that has evolved up to that point is used as part of the input state for the DRL algorithm. 
All in all, the state space was composed of the following components:
\begin{itemize}
    \item \emph{Intermediate maze of preceding step.} The maze generated by the previous step is encoded into a 2-dimensional grid map. Exercise rooms, as well as corridors and crossings, are mapped to predefined numerical values and given to the network as a 2-dimensional array, allowing the model architecture to make use of spatial information.
    \item \emph{Maze difficulty of preceding step.} As the goal of the algorithm is to build mazes with a certain difficulty, we found that explicitly feeding the difficulty of the maze that was generated in the preceding step enhanced the performance of the model. The difficulty is assessed by running a user simulation on the intermediate maze, which will be explained in more detail later.
    \item \emph{Number of crossings.} As the number of crossings that occur in a maze is one of the key factors to different levels of difficulty, we decided to directly feed the number of crossings of the previously generated maze into the network.
    \item \emph{Occupied exercise rooms.}
    The 2-dimensional grid of the intermediate maze only implicitly contains the information which exercise rooms are already occupied. Thus, we include the occupied rooms, encoded as one-hot vectors, in the state space to ease training. 
\end{itemize}
The 2D representation of the intermediate maze of the preceding step is fed into a block of convolutional layers. The resulting output is concatenated with all other components of the input and fed into a succeeding block of fully connected layers. 
The network is trained using the \emph{Deep Q-Learning} algorithm as proposed by Mnih et al. \cite{mnih2015human}. 

One crucial factor for the success of a DRL approach is the amount of training data. As human players need a certain amount of time for traversing each maze, it is not feasible to train the network solely with training data produced by real human players. 
Thus, we decided to pretrain our network on a user simulation that estimates the difficulty of a given maze.
This simulation is also used to estimate the difficulty of the intermediate mazes contained in the input states.
The user simulation is implemented to replicate the user's behavior as realistically as possible.
Thus, the simulation consists of an agent that has to find its way through a maze, whereas exercise rooms that are traversed add up to the physical effort demanded by the maze.
When the agent reaches a crossing the first time, the simulation randomly decides which path to take.
This approximation was chosen since players never see the maze from above and therefore have to choose paths randomly at the beginning of the game until they explore more of the maze.
However, as real users will probably be less likely to take the same path twice, the simulation agent remembers which way it has chosen at a certain crossing, and the probability of taking that path a second time when repeatedly passing the respective crossing is decreased.
To approximate the users' physical effort, we assigned an effort level to each exercise room, modeling the physical effort that has to be invested when passing the room.
For the final effort estimation of the whole maze, the effort levels of all passed exercise rooms are summed up until the end room is reached.

%% file: evaluation.tex
\section{Evaluation}

%% needs to be defined here since it is flushed to the next page
\begin{figure*}
    \centering
    \begin{minipage}{0.41\linewidth}
        \centering
        \includegraphics[width=0.75\linewidth]{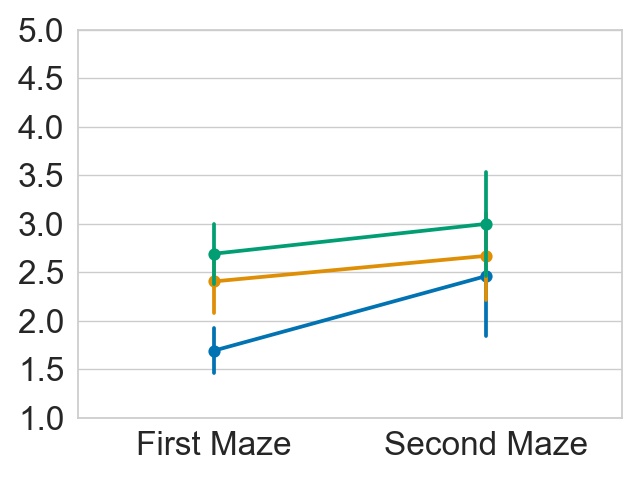}
    \end{minipage}
    \begin{minipage}{0.16\linewidth}
        \centering
        Measured Variable:
        \includegraphics[width= 0.9\linewidth]{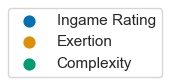}
    \end{minipage}
    \begin{minipage}{0.41\linewidth}
        \centering
        \includegraphics[width=0.75\linewidth]{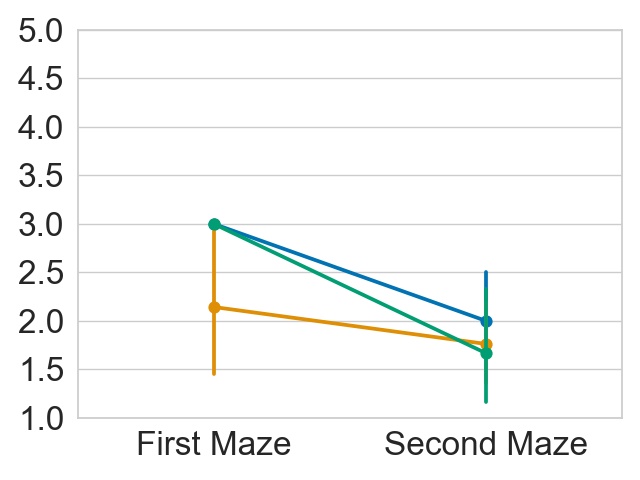}
    \end{minipage}
    \caption{The subjective ratings of the participants who were not satisfied with the first maze (left) and participants who were satisfied (right). 
    To unify the values, we linearly mapped the Borg scale (6-20) to a 5-point Likert scale  (1-5).
    Then we inverted the values for participants, who rated the first maze as too difficult such that an upwards trend indicates correct adaptation. 
    The Error bar shows the 95\% CI.}
    \label{fig:subjective_combined}
\end{figure*}

As an initial proof-of-concept, we conducted an exploratory human user study to verify whether our prototype is able to adjust the difficulty of the second of two consecutive mazes based on user feedback for the first maze.

\subsection{Experiment Design}
\paragraph{Research Questions}
For this study we had two main research questions: 1) is our approach able to adjust the difficulty for users who did not like the difficulty of the first maze (e.g., lower the difficulty for someone who found the first maze to be too hard and increase the difficulty for users who found it too easy.) 
and 2) can our approach sustain the difficulty for users that were satisfied with the difficulty of the first maze (e.g., rated the difficulty with 3 out of 5)?
For both questions, a key challenge is that the approach should not adjust the maze's complexity or the required physical exertion individually but keep the balance between those two aspects.

\paragraph{Methodology}
For a subjective evaluation of the difficulty, we recorded the participants' \emph{in-game rating} of the difficulty at the end of each maze, which is described at the beginning of section \ref{sec:approach} and which is also used as input for our DDA algorithm. 
In addition, we asked the participants how \emph{complex} and \emph{exerting} each maze was after they finished playing.
For complexity, we used a 5-point Likert scale (1-``not at all'' to 5-``extremely'').
For exertion, we used the Borg RPE (ratings for perceived exertion) scale proposed by Borg \cite{borg2006comparison}, which measures the perceived exertion on a range from 6 ("No exertion at all") to 20 ("Maximal exertion").

Furthermore, we recorded an electrocardiographic (ECG) signal to objectively measure the participants' exertion level during the VR session.
The ECG sensor was a 1-Lead sensor attached to the right side of the participants' upper body\footnote{The sensor was manufactured by Plux: https://www.biosignalsplux.com}.
The sensor was connected to an $8$-channel wireless hub, including eight generic inputs and one ground.
The operational sample rate was $1$kHz (i.e., $1000$ samples were recorded for every second).
The heart rate (HR) of the participants was calculated from the raw ECG signal, using the Python library Biosppy\footnote{https://github.com/PIA-Group/BioSPPy}.
Before calculating the HR, the ECG signal was filtered using Finite Impulse Response with bandpass frequency between 3 and 45 Hz.

In order to measure the participants' flow and their general satisfaction with our prototype, we used the game experience questionnaire (GEQ) \cite{IJsselsteij2013GEQ}.
The core module of this questionnaire was recently empirically evaluated by Law et al. \cite{law2018GEQ} and Johnson et al. \cite{johnson2018GEQ}.
Following their suggestions, we only used the categories Competence, Immersion, Flow, and Positive Affect and excluded the ”It was aesthetically pleasing” question from Immersion.

\paragraph{Procedure}
Before starting the experiment, the participants had to sign a consent form, followed by a short introduction to the setup and the procedure.
After that, the participants filled out a pre-questionnaire containing socio-demographic questions.
The pre-questionnaire additionally included two items about their previous experience with gaming (``I play games daily'') and VR (``I have experience with VR'') measured on a 5-point Likert scale (1-``strongly disagree'' to 5-``strongly agree'').
Then, they put on the HMD and the ECG sensor was attached to their body.
The VR session started with a tutorial level, which explained the controls of the game.
During this tutorial, a supervisor answered all questions which the participants might have about the controls.
After the tutorial, the participants were left to play the game without additional help from the supervisor, apart from warnings about hitting objects in the real world and clarifying the in-game ratings (some people thought a 5 would mean that they liked the maze, which we wanted to avoid.) 
Between the two mazes, there was a short break of approximately two minutes to bring participants' heart rates to a resting rate. 
Immediately after completing the second maze, the participants filled out a post-questionnaire that consisted of the items regarding the complexity and exertion of each maze and the GEQ.

\subsection{Results}
\paragraph{Participants} In order to test our proof-of-concept, we recruited 19 (5 female, 14 male) students with a mean age of $26.84$  (\textit{SD}=$4.55$).
Most participants had either a bachelor's or master's degree. 
Four only had a high school degree and one already possessed a doctoral degree.
On average, the participants reported a neutral gaming frequency ($M=2.79$), and an above-average VR experience ($M=3.68$).
Only four of the participants had never used HMDs before.

\paragraph{Research Question 1}

\begin{figure}
    \centering
    \includegraphics[width=0.75\linewidth]{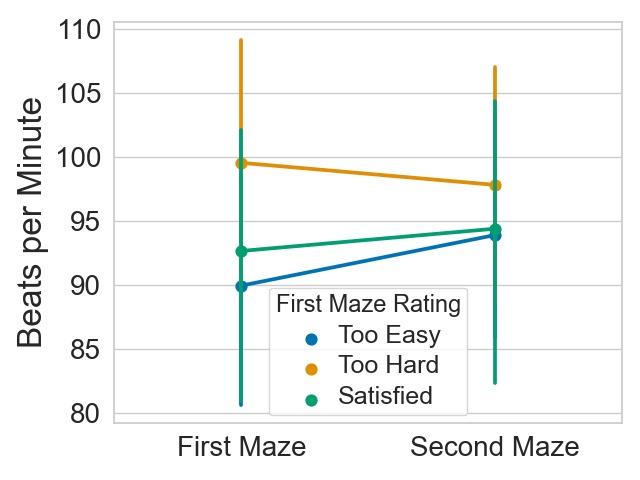}
    \caption{The mean HR measured in beats per minute for participants who found the first maze too easy, were satisfied with it or found it too difficult. Error bars show the 95\% Confidence Interval (CI).}
    \label{fig:heart_rate}
\end{figure}

For our first research question, we looked at the participants who were unsatisfied with the first maze.
In total, the first maze was too easy (rated 1 or 2) for 9 players and too hard (rated 4 or 5) for 4 players.
To unify those 13 players, we inverted the complexity and in-game ratings for the players who found it too hard (i.e. mapped 5 to 1 and 4 to 2).
The results for this unified \emph{adaption group} are shown on the left side of Fig. \ref{fig:subjective_combined}.
For the first maze, this \emph{adaption group} had an average in-game rating of $1.69$ (\textit{SD}=$0.46$) and for the second maze $2.46$ (\textit{SD}=$1.15$).
The subjective complexity rating went from a mean of $2.69$ (\textit{SD}=$0.61$) to $3.0$ (\textit{SD}=$0.96$).
Since the  Borg RPE scale is more nuanced, we did not unify it.
The exertion rating for participants who rated the first maze as too easy went from a mean of $10.0$ (\textit{SD}=$1.75$) to $10.44$ (\textit{SD}=$2.15$).
For participants who found the first maze too hard, the exertion rating went from $13.0$ (\textit{SD}=$1.41$) to $11.0$ (\textit{SD}=$2.0$).
The ECG signal results, namely the HR measured in beats per minute (bpm), are shown in Fig. \ref{fig:heart_rate}.
Since HR is a continuous variable and has no specific desired middle value, we did not combine the participants for this value.
The mean HR of participants who rated the first maze as too easy increased from $89.96$ bpm (\textit{SD}=15.05) to $93.90$ bpm (\textit{SD}=14.50).
For the participants that rated the first maze as too hard, the mean HR decreased from $99.53$ bpm (\textit{SD}=10.07) to $97.81$ bpm (\textit{SD}=9.68).

The trends above indicate that our prototype was able to adjust the difficultly for participants' who were unsatisfied with the first maze.

\paragraph{Research Question 2}

For our second research question, we only looked at participants who were satisfied with the first maze and gave it an in-game rating of 3.
The subjective results of this \emph{sustain group} are shown on the right side of Fig. \ref{fig:subjective_combined}.
Here, the mean in-game rating for the second maze was $2.0$ (\textit{SD}=0.58).
The subjectively reported exertion level decreased from a mean of $10.0$ (\textit{SD}=3.25) to $8.67$ (\textit{SD}=1.86) and the reported complexity decreased from $3.0$ (\textit{SD}=0.0) to $1.67$ (\textit{SD}=0.75).
The mean HR rose from $92.66$ bpm (\textit{SD}=12.39) to $94.40$ bpm (\textit{SD}=13.71), between the two mazes (Fig. \ref{fig:heart_rate} middle).

Here, the results are conflicted.
The subjective ratings indicate a decrease in difficulty while the HR suggests a slight increase in exertion.

\begin{figure}
    \centering
    \includegraphics[width=0.75\linewidth]{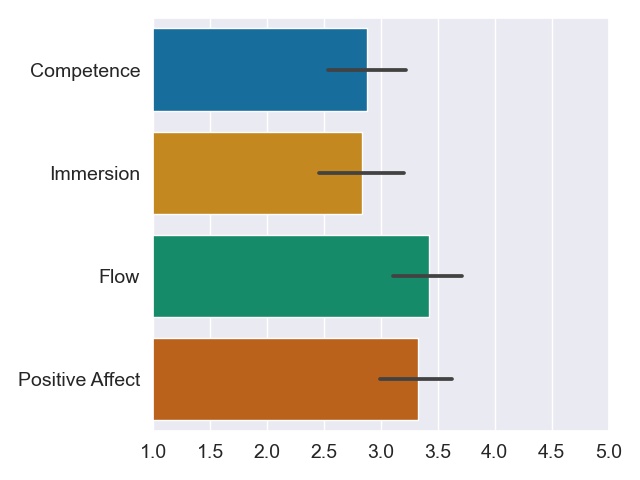}
    \caption{The results of the GEQ measuring the participants' flow and general satisfaction with our prototype. 
    The error bars show the 95\% CI.}
    \label{fig:game_exp_questionnaire}
\end{figure}

\paragraph{Game Experience}
The results of the GEQ are shown in Fig. \ref{fig:game_exp_questionnaire}.
Noticeably, the flow value is above average with a mean of $3.42$ (\textit{SD}=$0.71$).

%% file: discussion.tex
\section{Discussion}

The results of our exploratory user study indicate that there is potential in procedurally generating levels with a fitting difficulty for DDA in VR exergames.
Our prototype was able to adapt both the cognitive as well as physical difficulty of the second maze according to the needs of participants who were unsatisfied with the first maze.
For people who were already satisfied with the first maze, the results are conflicted.
The subjective responses indicate that the second maze was easier while the HR shows slightly increased exertion.
Based on the positive results for participants who were unsatisfied with the first maze, we think that a third maze would positively adapt to the participants again. 
Across all groups, the heart rate results for the second maze are promising since they are comparable to the approximately 95 bpm that Charoensook et al. \cite{charoensook2019} measured for the two most exerting VR games during their study on exertion during VR gameplay.   
The results of the GEQ (see Fig. \ref{fig:game_exp_questionnaire}) show that participants experienced an above-average level of flow even though our prototype received below-average ratings for competence and immersion. 
This suggests that the DDA algorithm, which was the focus of our interest, helped to keep participants in a state of flow, even though other aspects of the game were not developed to their full potential.

\subsection{Study Limitations}
While the results of our evaluation are promising, they should only be taken as a proof-of-concept.
The goal of our exploratory study was not to test a finished exergame but to test the plausibility of combining DDA and PCG to create VR exergame levels that match the participants' capabilities.

Because of this exploratory nature of our study and the high associated effort and time investment for participants, we only obtained a relatively small number of participants.
While this is common among exploratory evaluations of DDA in exergames (\cite{burke09stroke_rehab,hocine2011}), a bigger number of participants would be needed to statistically verify the usability of a finished VR exergame that uses our proposed approach.
Furthermore, in order to see the full adaptation capabilities of such a finished exergame, it would need to be evaluated over a longer period of time in which participants use the exergame regularly.
Based on the proof-of-concept in this work, a full study with a more developed exergame could investigate the potential of the approach in more detail.

\subsection{Lessons Learned}
To inform the future design of adaptive level generation systems for exergames, we want to conclude the discussion with some lessons we learned during the user study with our prototype.

\paragraph{Adapt the simulation} The user simulation in our prototype is static since it is only used for pretraining the model and to get an estimate of the difficulty of the intermediate mazes. 
The final adaptation to the player is left to the DRL algorithm that generates the maze.
In retrospect, we think that adapting the simulation to better reflect the player, similar to \cite{wheat2015}, might speed up the adaption.
In particular, we think that this might have prevented the perceived drop in difficulty for participants that were satisfied with the first maze since it could account for players getting used to the game.

\paragraph{Keep levels short} 
We tried to create short mazes with less than 8 rooms in most mazes.
However, physically doing the exercises took longer than we anticipated.
The participants in our study spent up to 10 minutes in each maze. 
This drastically reduces the number of learning steps that can be done in a given amount of time and therefore slows down adaptation. 
In the future, we will aim to create even shorter levels.
For instance, we would like to explore other methods for creating variants of our exercise rooms with different difficulty, instead of only increasing the repetitions.

%% file: conclusion.tex
\section{Conclusion}

In this paper, we show a first prototype of how procedural generation can be incorporated into the DDA system of a VR exergame using deep reinforcement learning.
The results of our exploratory user study are promising and indicate that the generated levels can indeed adapt both the required physical and cognitive effort according to the player's capabilities.
However, our particular prototype showed some problems that could guide the design of future systems.
Two main problems were that the generated levels took too long to allow for fast adaption and that we used a static user simulation.

Based on our results, we are confident that procedural generation of levels with appropriate difficulty, in addition to traditional in-game parameter adjustment, can improve future DDA systems for exergames.

%% file: main.bbl
\begin{thebibliography}{10}
\providecommand{\url}[1]{#1}
\csname url@samestyle\endcsname
\providecommand{\newblock}{\relax}
\providecommand{\bibinfo}[2]{#2}
\providecommand{\BIBentrySTDinterwordspacing}{\spaceskip=0pt\relax}
\providecommand{\BIBentryALTinterwordstretchfactor}{4}
\providecommand{\BIBentryALTinterwordspacing}{\spaceskip=\fontdimen2\font plus
\BIBentryALTinterwordstretchfactor\fontdimen3\font minus
  \fontdimen4\font\relax}
\providecommand{\BIBforeignlanguage}[2]{{%
\expandafter\ifx\csname l@#1\endcsname\relax
\typeout{** WARNING: IEEEtran.bst: No hyphenation pattern has been}%
\typeout{** loaded for the language `#1'. Using the pattern for}%
\typeout{** the default language instead.}%
\else
\language=\csname l@#1\endcsname
\fi
#2}}
\providecommand{\BIBdecl}{\relax}
\BIBdecl

\bibitem{yoo2020}
S.~Yoo, P.~Gough, and J.~Kay, ``Embedding a {VR} game studio in a sedentary
  workplace: Use, experience and exercise benefits,'' in \emph{{CHI} '20: {CHI}
  Conference on Human Factors in Computing Systems}.\hskip 1em plus 0.5em minus
  0.4em\relax {ACM}, 2020, pp. 1--14.

\bibitem{costa2019virtual}
M.~T.~S. Costa, L.~P. Vieira, E.~de~Oliveira~Barbosa, L.~M. Oliveira,
  P.~Maillot, C.~A.~O. Vaghetti, M.~G. Carta, S.~Machado, V.~Gatica-Rojas, and
  R.~S. Monteiro-Junior, ``Virtual reality-based exercise with exergames as
  medicine in different contexts: A short review,'' \emph{Clinical practice and
  epidemiology in mental health: CP \& EMH}, vol.~15, p.~15, 2019.

\bibitem{haskell2007physical}
W.~L. Haskell, I.-M. Lee, R.~R. Pate, K.~E. Powell, S.~N. Blair, B.~A.
  Franklin, C.~A. Macera, G.~W. Heath, P.~D. Thompson, and A.~Bauman,
  ``Physical activity and public health: updated recommendation for adults from
  the american college of sports medicine and the american heart association,''
  \emph{Circulation}, vol. 116, no.~9, p. 1081, 2007.

\bibitem{czikszentmihalyi1990flow}
M.~Czikszentmihalyi, \emph{Flow: The psychology of optimal experience}.\hskip
  1em plus 0.5em minus 0.4em\relax New York: Harper \& Row, 1990.

\bibitem{csikszentmihalyi2014toward}
M.~Csikszentmihalyi, ``Toward a psychology of optimal experience,'' in
  \emph{Flow and the foundations of positive psychology}.\hskip 1em plus 0.5em
  minus 0.4em\relax Springer, 2014, pp. 209--226.

\bibitem{Yoo2017}
S.~Yoo, C.~Parker, and J.~Kay, ``Designing a personalized {VR} exergame,'' in
  \emph{Adjunct Publication of the 25th Conference on User Modeling, Adaptation
  and Personalization, {UMAP}}.\hskip 1em plus 0.5em minus 0.4em\relax {ACM},
  2017, pp. 431--435.

\bibitem{pezzera2019}
M.~Pezzera, A.~Tironi, J.~Essenziale, R.~Mainetti, and N.~A. Borghese,
  ``Approaches for increasing patient's engagement and motivation in
  exer-games-based autonomous telerehabilitation,'' in \emph{7th {IEEE}
  International Conference on Serious Games and Applications for Health,
  SeGAH}, 2019, pp. 1--8.

\bibitem{finkelstein2011}
S.~L. Finkelstein, A.~Nickel, Z.~Lipps, T.~Barnes, Z.~Wartell, and E.~A. Suma,
  ``Astrojumper: Motivating exercise with an immersive virtual reality
  exergame,'' \emph{Presence Teleoperators Virtual Environ.}, vol.~20, no.~1,
  pp. 78--92, 2011.

\bibitem{charoensook2019}
T.~Charoensook, M.~Barlow, and E.~Lakshika, ``Heart rate and breathing
  variability for virtual reality game play,'' in \emph{2019 IEEE 7th
  International Conference on Serious Games and Applications for Health
  (SeGAH)}, 2019, pp. 1--7.

\bibitem{hunicke2005}
R.~Hunicke, ``The case for dynamic difficulty adjustment in games,'' in
  \emph{Proceedings of the International Conference on Advances in Computer
  Entertainment Technology, {ACE}}, 2005, pp. 429--433.

\bibitem{zohaib2018}
M.~Zohaib, ``Dynamic difficulty adjustment {(DDA)} in computer games: {A}
  review,'' \emph{Adv. Hum. Comput. Interact.}, vol. 2018, pp.
  5\,681\,652:1--5\,681\,652:12, 2018.

\bibitem{andrade2005}
G.~Andrade, G.~L. Ramalho, H.~Santana, and V.~Corruble, ``Challenge-sensitive
  action selection: an application to game balancing,'' in \emph{Proceedings of
  the 2005 {IEEE/WIC/ACM} International Conference on Intelligent Agent
  Technology}, 2005, pp. 194--200.

\bibitem{pagalyte2020}
E.~Pagalyte, M.~Mancini, and L.~Climent, ``Go with the flow: Reinforcement
  learning in turn-based battle video games,'' in \emph{{IVA} '20: {ACM}
  International Conference on Intelligent Virtual Agents}, 2020, pp.
  44:1--44:8.

\bibitem{sekhavat2017}
Y.~A. Sekhavat, ``{MPRL:} multiple-periodic reinforcement learning for
  difficulty adjustment in rehabilitation games,'' in \emph{5th {IEEE}
  International Conference on Serious Games and Applications for Health,
  SeGAH}.\hskip 1em plus 0.5em minus 0.4em\relax {IEEE} Computer Society, 2017,
  pp. 1--7.

\bibitem{yannakakis2018artificial}
G.~N. Yannakakis and J.~Togelius, \emph{{Artificial Intelligence and
  Games}}.\hskip 1em plus 0.5em minus 0.4em\relax Springer, 2018,
  \url{http://gameaibook.org}.

\bibitem{jennings2010polymorph}
M.~Jennings-Teats, G.~Smith, and N.~Wardrip-Fruin, ``Polymorph: dynamic
  difficulty adjustment through level generation,'' in \emph{Proceedings of the
  2010 Workshop on Procedural Content Generation in Games}, 2010, pp. 1--4.

\bibitem{shaker2010}
N.~Shaker, G.~N. Yannakakis, and J.~Togelius, ``Towards automatic personalized
  content generation for platform games,'' in \emph{Proceedings of the Sixth
  {AAAI} Conference on Artificial Intelligence and Interactive Digital
  Entertainment, {AIIDE}}, 2010.

\bibitem{gonzalez-duque2020}
M.~G. Duque, R.~B. Palm, D.~Ha, and S.~Risi, ``Finding game levels with the
  right difficulty in a few trials through intelligent trial-and-error,'' in
  \emph{{IEEE} Conference on Games, CoG}, 2020, pp. 503--510.

\bibitem{wheat2015}
D.~{Wheat}, M.~{Masek}, C.~P. {Lam}, and P.~{Hingston}, ``Dynamic difficulty
  adjustment in 2d platformers through agent-based procedural level
  generation,'' in \emph{2015 IEEE International Conference on Systems, Man,
  and Cybernetics}, 2015, pp. 2778--2785.

\bibitem{van2013designing}
R.~Van~der Linden, R.~Lopes, and R.~Bidarra, ``Designing procedurally generated
  levels,'' in \emph{Ninth Artificial Intelligence and Interactive Digital
  Entertainment Conference}, 2013.

\bibitem{burke09stroke_rehab}
J.~W. Burke, M.~D.~J. McNeill, D.~Charles, P.~J. Morrow, J.~Crosbie, and
  S.~McDonough, ``Optimising engagement for stroke rehabilitation using serious
  games,'' \emph{Vis. Comput.}, vol.~25, no.~12, pp. 1085--1099, 2009.

\bibitem{burke2010designing}
J.~Burke, M.~McNeill, D.~Charles, P.~Morrow, J.~Crosbie, and S.~McDonough,
  ``Designing engaging, playable games for rehabilitation,'' in
  \emph{Proceedings of the 8th international conference on disability, virtual
  reality \& associated technologies}, 2010, pp. 195--201.

\bibitem{alankus2010}
G.~Alankus, A.~Lazar, M.~May, and C.~Kelleher, ``Towards customizable games for
  stroke rehabilitation,'' in \emph{Proceedings of the 28th International
  Conference on Human Factors in Computing Systems, {CHI}}, 2010, pp.
  2113--2122.

\bibitem{hocine2011}
N.~Hocine and A.~Goua{\"{\i}}ch, ``Therapeutic games' difficulty adaptation: An
  approach based on player's ability and motivation,'' in \emph{16th
  International Conference on Computer Games, {CGAMES}}, 2011, pp. 257--261.

\bibitem{smeddinck2013}
J.~D. Smeddinck, S.~Siegel, and M.~Herrlich, ``Adaptive difficulty in exergames
  for parkinson's disease patients,'' in \emph{Graphics Interface 2013, {GI}
  '13, Proceedings}, 2013, pp. 141--148.

\bibitem{pezzera2020dynamic}
M.~Pezzera and N.~A. Borghese, ``Dynamic difficulty adjustment in exer-games
  for rehabilitation: a mixed approach,'' in \emph{8th {IEEE} International
  Conference on Serious Games and Applications for Health, SeGAH}, 2020, pp.
  1--7.

\bibitem{andrade2016}
K.~O. Andrade, T.~B. Pasqual, G.~A.~P. Caurin, and M.~K. Crocomo, ``Dynamic
  difficulty adjustment with evolutionary algorithm in games for rehabilitation
  robotics,'' in \emph{2016 {IEEE} International Conference on Serious Games
  and Applications for Health, SeGAH}.\hskip 1em plus 0.5em minus 0.4em\relax
  {IEEE} Computer Society, 2016, pp. 1--8.

\bibitem{buskies2003ruckenfitness}
W.~Buskies and N.~Demski, \emph{R{\"u}ckenfitness: Grundlagen, {\"U}bungen,
  Spiele}.\hskip 1em plus 0.5em minus 0.4em\relax Limpert, 2003.

\bibitem{Khalifa2020}
A.~Khalifa, P.~Bontrager, S.~Earle, and J.~Togelius, ``Pcgrl: Procedural
  content generation via reinforcement learning,'' \emph{Proceedings of the
  AAAI Conference on Artificial Intelligence and Interactive Digital
  Entertainment}, vol.~16, no.~1, pp. 95--101, Oct. 2020.

\bibitem{mnih2015human}
V.~Mnih, K.~Kavukcuoglu, D.~Silver, A.~A. Rusu, J.~Veness, M.~G. Bellemare,
  A.~Graves, M.~Riedmiller, A.~K. Fidjeland, G.~Ostrovski \emph{et~al.},
  ``Human-level control through deep reinforcement learning,'' \emph{nature},
  vol. 518, no. 7540, pp. 529--533, 2015.

\bibitem{borg2006comparison}
E.~Borg and L.~Kaijser, ``A comparison between three rating scales for
  perceived exertion and two different work tests,'' \emph{Scandinavian journal
  of medicine \& science in sports}, vol.~16, no.~1, pp. 57--69, 2006.

\bibitem{IJsselsteij2013GEQ}
W.~IJsselsteijn, Y.~{de Kort}, and K.~Poels,
  \emph{\BIBforeignlanguage{English}{The Game Experience Questionnaire}}.\hskip
  1em plus 0.5em minus 0.4em\relax Technische Universiteit Eindhoven, 2013.

\bibitem{law2018GEQ}
E.~L. Law, F.~Br{\"{u}}hlmann, and E.~D. Mekler, ``Systematic review and
  validation of the game experience questionnaire {(GEQ)} - implications for
  citation and reporting practice,'' in \emph{The Annual Symposium on
  Computer-Human Interaction in Play, {CHI} {PLAY}}, 2018, pp. 257--270.

\bibitem{johnson2018GEQ}
D.~M. Johnson, M.~J. Gardner, and R.~Perry, ``Validation of two game experience
  scales: The player experience of need satisfaction {(PENS)} and game
  experience questionnaire {(GEQ)},'' \emph{Int. J. Hum. Comput. Stud.}, vol.
  118, pp. 38--46, 2018.

\end{thebibliography}
